\documentclass[]{aa}
\usepackage{epsfig}
\usepackage{graphics}
\usepackage{subfigure}
\usepackage{natbib}
\usepackage{amsmath}
\usepackage{enumerate}
\usepackage{threeparttable}
\usepackage{caption}
\usepackage{ulem}
\usepackage[usenames,dvipsnames,svgnames,table]{xcolor}

\begin{document}

\title{Mechanical feedback effects on primordial black hole accretion}

\author{V. Bosch-Ramon\inst{1} \and N. Bellomo\inst{1}}
  
\institute{Departament de F\'{i}sica Qu\`antica i Astrof\'{i}sica, Institut de Ci\`encies del Cosmos (ICC), Universitat de Barcelona (IEEC-UB), Mart\'{i} i Franqu\`es 1, E08028 Barcelona, Spain. \\\\  \email{vbosch@fqa.ub.edu, nicola.bellomo@icc.ub.edu}}


\titlerunning{Outflows and accretion in primordial black holes}

\abstract{Dark matter may consist, at least partially, of primordial black holes formed during the radiation-dominated era. The radiation produced by accretion onto primordial black holes leaves characteristic signatures on the properties of the medium at high redshift, before and after Hydrogen recombination. Therefore, reliable modeling of accretion onto these objects is required to obtain robust constraints on their abundance.}{We investigate the effect of mechanical feedback, that is, the impact of outflows (winds and -or- jets) on the medium, on primordial black hole accretion, and thereby on the associated radiation.}{Using analytical and numerical calculations, we studied for the first time the possibility that outflows can reduce the accretion rate of primordial black holes with masses similar to those detected by the LIGO-Virgo collaboration.}{Despite the complexity of the accretion rate evolution, mechanical feedback is useful in to significantly reducing the primordial black hole accretion rate, at least by one order of magnitude, when outflows are aligned with the motion of the compact object. If the outflow is perpendicular to the direction of motion, the effect is less important, but it is still non-negligible.}{Outflows from primordial black holes, even rather weak ones, can significantly decrease the accretion rate, effectively weakening abundance constraints on these objects. Our results motivate further numerical simulations with a more realistic setup, which would yield more precise quantitative predictions.}


\maketitle

\section{Introduction} 

The first gravitational waves detected by LIGO-Virgo were produced by the merger of two heavy stellar-mass black holes (BHs) \citep{lig16}. These detections revived the idea that primordial black holes \citep[PBHs;][]{zel67,haw71,car74,cha75} may be an important component of dark matter \citep[][]{bir16,sas16,cle16}, particularly with regard to those with masses around $10-100$~M$_\odot$. Among the different astrophysical and cosmological consequences of PBH existence \citep[see, e.g.,][and references therein]{sas18}, the radiation produced by PBH accretion could have left a detectable imprint on the cosmic microwave background \citep[CMB;][]{ric08,ali17,pou17,ber17,nak18} and the 21-cm Hydrogen signal \citep[][]{ber18,hek18,men19,hut19} and could be producing detectable radio and X-ray emission in our galaxy \citep[][]{gag17,man19}. 

The characterization of accreting PBHs is based on a number of assumptions that have not been fully tested yet. The robustness of PBH abundance constraints, coming from CMB, 21-cm signal, and galactic emission, depends significantly on those assumptions. In particular, one important aspect of PBH accretion-emission physics that has been rarely discussed in the literature concerns the indirect impact on accretion of outflows, winds and (or) jets, produced by the PBHs themselves. This effect, known as mechanical feedback, consists of a reduction of gas accretion caused by outflows sweeping the medium away, as this leaves only diluted and hot gas to accrete. Mechanical feedback is, in fact, an important astrophysical process in general, and has been discussed for instance in the context of heavy stellar-mass BHs as those detected by LIGO \citep[][]{iok17}, supermassive BHs growth \citep[see, e.g.,][and references therein]{lev18,zei19}, and other astrophysical objects such as stars, galaxies and clusters \citep[e.g.,][]{sok16,gru19,li19}. 

The aim of the present work is to estimate for the first time, using analytical and numerical tools, the potential reduction of the accretion rate, hence in the associated radiation, caused by the production of outflows in a moving PBH. \cite{li19} already tentatively explored, through a numerical study, the impact of mechanical feedback on Bondi-Hoyle-Lyttleton accretion in different astrophysical contexts. Here we consider the PBH case due to its far-reaching astrophysical and cosmological consequences. Thus, we focus on the case in which a PBH moves with a supersonic speed through an almost homogeneous medium, similar to that of the Universe at high redshift, when non-linear structures had not yet formed, both before and after Hydrogen recombination. Therefore the results of this work can be applied both to existing CMB constraints and to future 21-cm constraints, but they also hold relevance for PBH (and isolated BHs in general) detectability in the present-day Universe. 

The paper is structured as follows: first we provide an analytical characterization of the impact of outflows on the accretion process (Sections~\ref{sec:basics_accretion_ejection} and~\ref{sec:mechanical_feedback_analytic}). Then, we present the results of numerical simulations to complement the analytical characterization (Sect.~\ref{sec:mechanical_feedback_simulations}). Finally, we conclude with a summary and a discussion of our findings (Sect.~\ref{sec:conclusions}). Throughout this work, we adopt the convention $Q_{b} = Q / 10^b$, where $Q$ is any physical quantity measured in cgs units\footnote{Useful conversion factors between different systems of units: $1\ {\rm g\, s^{-1}} \approx 10^{-26}\ {\rm M}_\odot\, {\rm yr^{-1}} \approx 10^{21}\, {\rm erg\, s^{-1}}/ c^{2}$, where the speed of light~$c$ in cgs units is~$c^2 \approx 10^{21}\ {\rm cm^2\, s^{-2}}$; the gravitational constant in cgs units is~$G \approx 6.7\times 10^{-8}\ {\rm cm^3\, g^{-1}\, s^{-2}}$; $1\, {\rm cm} \approx 3 \times 10^{-19}\, {\rm pc}$; and $1\, {\rm s} \approx 3 \times 10^{-8} \, {\rm yr}$.}, unless indicated otherwise. Also, the symbol $\sim$ indicates an order of magnitude estimate, whereas the symbol $\approx$ is used when we provide an approximate numerical estimate of a given quantity.


\section{Basics of accretion and outflow physics}
\label{sec:basics_accretion_ejection}

\subsection{Accretion}\label{accre}

We consider a PBH of mass $M_{\rm PBH}$ that moves with supersonic speed $v_{\rm PBH}$ through an almost homogeneous medium with density $\rho_{\rm m} = \left\langle m \right\rangle n_{\rm m}$, number density~$n_{\rm m}$, and average particle mass $\left\langle m \right\rangle \sim m_H \approx 1.7\times 10^{-24}\ {\rm g}$ (i.e., the Hydrogen atom mass). We assume as reference case a PBH with mass $M_{\rm PBH}=30$~M$_\odot$ moving with a speed of~$v_{\rm PBH}=3\times 10^6\ {\rm cm\ s^{-1}}$ in a medium with number density $n_{\rm m} = 1\ {\rm cm^{-3}}$ (i.e., $M_{\rm PBH,1.5}\approx 1$, $v_{\rm PBH,6.5}\approx 1$, $n_{\rm m,0} = 1$).

As proposed in \cite{ali17}, if PBHs are all the dark matter, their typical speed is approximately constant before Hydrogen recombination, with $v_\mathrm{PBH}\approx 3\times 10^6\ \mathrm{cm\ s^{-1}}$ \citep[][]{dvo14}, and $v_\mathrm{PBH}\propto (1+z)$ afterwards \citep[][]{tse10}, where $z$ is the redshift. Since PBHs move supersonically with respect to baryons until relatively low redshift ($z\approx 20$), our set-up roughly captures both pre- and (early) post-recombination scenarios. The reference value taken for the number density of the medium, $n_{\rm m}=1\ \mathrm{cm}^{-3}$, corresponds to the baryon number density $n_{\rm b}(z) = n_{\rm b0} (1+z)^3$ at redshift $z \approx 150$, where $n_{\rm b0} \sim \Omega_{\rm b0} \rho_{\rm 0c} / m_H \approx 3\times 10^{-7}\ {\rm cm^{-3}}$ is the present-day baryon number density, $\Omega_{\rm b0}\approx 0.05$ the baryon fractional abundance at present time, and $\rho_{\rm 0c} \approx 10^{-29}\ {\rm g\ cm^{-3}}$ the present-day critical density.\footnote{We note that, in the adiabatic case of mechanical feedback, our conclusions will not depend on $n_{\rm m}$, and thus its actual value is not of much relevance here.}

A supersonic PBH accretes approximately at the Bondi-Hoyle-Lyttleton accretion rate \citep[][]{hoy39, bon44}: 
\begin{equation}
\begin{aligned}
\dot{M}_{\rm PBH} &\approx \pi r_{\rm acc}^2 \rho_{\rm m} v_{\rm PBH} \approx
10^{13} M_{\rm PBH,1.5}^2\, n_{\rm m,0}\, v_{\rm PBH,6.5}^{-3} \ {\rm g\, s^{-1}}\\&\approx 
1.6\times 10^{-13}M_{\rm PBH,1.5}^2\, n_{\rm m,0}\, v_{\rm PBH,6.5}^{-3} \ {\rm M_\odot\, yr^{-1}}
,
\end{aligned}
\label{eq:bondi_hoyle_accretion_rate}
\end{equation}
where the effective accretion radius $r_\mathrm{acc}$, which defines the PBH sphere of influence radius, is given by:
\begin{equation}
r_{\rm acc} \sim 2\, G M_{\rm PBH} v_{\rm PBH}^{-2} \approx 8 \times 10^{14} M_{\rm PBH,1.5}\, v_{\rm PBH,6.5}^{-2} \ {\rm cm}.
\label{eq:accretion_radius}
\end{equation}

In the case where the PBH moves with subsonic speed, we have to substitute $v_{\rm PBH}$ in Eq.~\eqref{eq:accretion_radius} with the sound speed of the medium \citep{bon52}. In that situation, accretion would be closer to the spherical case, at least on scales $\lesssim r_{\rm acc}$ \citep[but not necessarily in the close vicinity of the PBH;][]{wat19}. This would be the case, for instance, at redshift $z\lesssim 20$, but we do not investigate that epoch here due to its high complexity, as it coincides with the formation of the first structures.

Processes in the PBH vicinity may reduce the accretion rate in that region. Typically, they are accounted for by defining the real accretion rate as $\dot{M} = \lambda \dot{M}_{\rm PBH}$, where $\dot{M}_{\rm PBH}$ is the accretion rate defined in Eq.~\eqref{eq:bondi_hoyle_accretion_rate}, and the parameter $\lambda \lesssim 1$ encloses the physical effects of pressure, viscosity, radiation feedback and other processes on scales close to the PBH gravitational radius ($r_{\rm g}=GM_{\rm PBH}/c^2\approx 5\times 10^6$~cm for $M_{\rm PBH}=30\,$M$_\odot$). In fact, $\lambda$ captures the presence of gravitationally unbounded flows of any type. In particular, the production of fast winds or jets can significantly reduce the accretion rate on those scales \citep[i.e., $\lambda\ll 1$; see, e.g.,][in the context of isolated stellar-mass and supermassive BHs]{tsu18,bu19}. Here, since we are interested only on mechanical feedback on scales comparable to $r_{\rm acc}$, any process that reverts accretion close to the PBH is phenomenologically treated as outflow on larger scales. Thus, our reference value for the accretion rate is $\dot{M}_{\rm PBH}$, that is, we take $\lambda=1$.

\begin{figure}        
\includegraphics[width=\columnwidth]{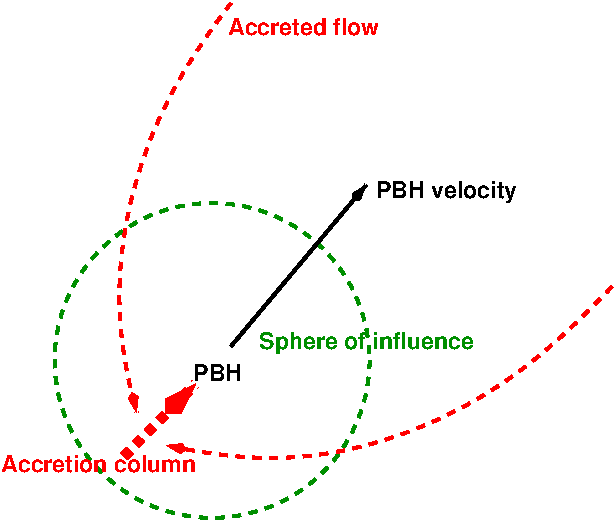}
\caption{Sketch of the no-outflow case, where the PBH is moving supersonically through a medium, and it is accreting material from its sphere of influence through a gas column forming opposite to the direction of motion. Sketch not to scale.}
\label{fig:sketch_column_accretion}
\end{figure}

Relatively far from the PBH, accretion takes place through a column that forms behind the
accretor, as schematically drawn in Fig.~\ref{fig:sketch_column_accretion}. In the case of a
perfectly homogeneous medium, accretion is axisymmetric, and matter is channeled by this
gravitationally bounded gas column straight to the accretor. However, it is expected that
small perturbations, which would be certainly present in a cosmological context, would lead to a break of the
axisymmetry of the accretion process, which can be then enhanced by instability growth. This
can lead to the accretion of some angular momentum (see, e.g., \citealt{fog05,lor15}; see also
\citealt{wat19} for the spherical case). In fact, density inhomogeneities accreted with
different impact parameters will already provide some angular momentum. In these contexts,
even if the average angular momentum accreted in the long-term is very small, it is possible
to form a short-lived small disk close to the PBH \citep[see, e.g.,][and references therein]{ago02}. 

To further characterize accretion in the vicinity of the PBH, it is informative to compare the BH accretion luminosity to the Eddington luminosity. Accounting only for Thomson scattering on ionized Hydrogen, the Eddington luminosity is given by $L_{\rm Edd} \approx 3.8\times 10^{39}\, M_{\rm PBH, 1.5}\ {\rm erg\, s^{-1}}$. Defining here the Eddington accretion rate as $\dot{M}_{\rm Edd}=10 L_{\rm Edd} / c^2$, we obtain the ratio of the PBH accretion rate to the Eddington rate given by:
\begin{equation}
\frac{\dot{M}_{\rm PBH}}{\dot{M}_{\rm Edd}} \approx 2.5 \times 10^{-7} M_{\rm PBH,1.5}\, n_{\rm m,0}\, v_{\rm PBH,6.5}^{-3}\,,
\end{equation}
which shows that, for a very broad range of parameter values, $\dot{M}_{\rm PBH} / \dot{M}_{\rm Edd}\ll 0.1$. Such a low Eddington accretion fraction implies that in our scenario the accreted gas will not cool efficiently on scales $\gtrsim r_{\rm g}$, leading to the formation of a faint and hot (virialized) accretion structure close to the PBH \citep[see, e.g.,][]{sha73}. As we will see, such a configuration is prone to outflow formation.

Observational cosmology is unable to establish whether inhomogeneities lead to net angular momentum
accretion because of the small spatial scales involved. The key quantities that regulate angular
momentum accretion are amplitude and spatial scale of the inhomogeneities. Regarding the former, we
note that the development of instabilities in the inner accretion structure can help to amplify rather
smooth inhomogeneities. Concerning the latter, by characterizing the inhomogeneity scale as $l_{\rm
inh}$, one can estimate the duration of net angular momentum accretion: $t_l\sim l_{\rm inh}/v_{\rm
PBH}$. For inhomogeneities to be important, it is required that: (i) $t_l \gg r_{\rm g}/c$, that is, $t_l$
is larger than the wind formation time scale; and (ii) $t_l \lesssim r_{\rm acc}/v_{\rm PBH}$, that is,
$t_l$ is smaller than the PBH sphere of influence crossing time. Otherwise, (i) the wind will not have
time to form, or (ii) the gradients may be too small to accrete a non-negligible amount of net angular
momentum. Given the random nature of the inhomogeneities, both in amplitude and in $l_{\rm ihn}$, it is
expected that the outflow will change orientation on a time $\sim t_l$. Given that the evolution
timescale of the whole system is expected to be longer than any relevant $t_l$ (see below), the outflow
could effectively act, on average, as a very broad wind at large scales.

\subsection{Outflows}

A bipolar outflow (wind and -or- jet) from an accreting PBH can affect its own energetics, and the radiation from accretion, by sweeping away medium gas on scales $\sim r_{\rm acc}$. Nevertheless, the launch of such an outflow requires: (i) a strong magnetic field of specific geometry; and (or) (ii) an excess of thermal energy in some of the accreted gas, that is, positive energy gas that turns into a wind. 

Regarding (i), we must note first that the magnetic field in the accreted primordial gas is expected to be very small, although the Biermann battery mechanism may alleviate this problem in the accretion flow, and the magnetorotational instability could produce an enhancement of magnetization and turbulence, thereby generating the viscosity needed for accretion \cite[see, e.g.,][]{saf18}\footnote{It is worth mentioning that a magnetized accretion disk with low viscosity can also launch outflows \citep{stas10}.}. However, the magnetic field seed generated by the Biermann battery is toroidal, whereas jet launching requires some vertical magnetic flux 
\citep[see, e.g., the discussion in][]{sot19}. Nevertheless, if enough vertical flux is present close to the PBH, the inner accretion region can efficiently launch a collimated, supersonic bipolar outflow \citep{bla82}. In addition, if the PBH is spinning fast, relativistic, highly collimated jets can be also produced in the PBH ergosphere through the Blandford-Znajek \citep{bla77} mechanism (even if accretion is quasi-spherical, as discussed, for example, in \citealt{bar12a,bar12b}). 

In the case of (ii), regardless of the presence of jets, a weakly accreting, geometrically thick disk can form winds through thermal pressure gradients \citep[see, e.g.,][where jets are also shown to form if the BH is spinning]{sad16}. Such (less collimated) outflows, produced by an excess of thermal energy in the accretion structure, are less dependent on the details of the magnetic field. Thus, to be conservative, hereafter we assume that the outflow has a thermal origin, and carries only a small fraction $\epsilon$ of the rate of accreted energy  $\sim \dot{M}_{\rm PBH}c^2$. The power of each constituent of the bipolar supersonic outflow can be characterized as:
\begin{equation}
L_{\rm w} =\epsilon \dot{M}_{\rm PBH} c^2 \approx 10^{30} \epsilon_{-4} M_{\rm PBH,1.5}^2 n_{\rm m,0} v_{\rm PBH,6.5}^{-3}\ {\rm erg\, s^{-1}}.
\label{eq:outflow_luminosity}
\end{equation}
Alternatively, we can write
\begin{equation}
L_{\rm w}=(\dot{M}_{\rm ejec}/2)v_{\rm w}^2/2\,
\label{lonr}
\end{equation}
or
\begin{equation}
L_{\rm w}=\gamma_{\rm w}(\dot{M}_{\rm ejec}/2)c^2
\label{lor}
\end{equation}
in the non-relativistic and the highly relativistic (cold outflow) case, respectively, where $v_{\rm w}$ is the outflow speed, $\dot{M}_{\rm ejec}$ the total ejected mass, and $\gamma_{\rm w}$ the outflow Lorentz factor. Equations (\ref{eq:outflow_luminosity}) and (\ref{lonr}) imply
\begin{equation}
\dot{M}_{\rm ejec}\approx 0.36\,\epsilon_{-4}\,v_{\rm w,9}^{-2}\dot{M}_{\rm PBH}\,,
\label{ejec}
\end{equation}
which shows that $v_{\rm w}\gtrsim 6\times 10^8\epsilon_{-4}^{1/2}$~cm~s$^{-1}$ to ensure that $\dot{M}_{\rm ejec}<\dot{M}_{\rm PBH}$ (i.e., large $\epsilon$ values require highly relativistic outflows). 

We can also estimate a minimum $\epsilon$ for the outflow to escape the vicinity of the PBH: The outflow propagation should proceed roughly unimpeded if accretion is rather equatorial, away from the outflow close to the PBH. However, beyond some distance, hard to ascertain but prescribed here by the parametric length $h_{\rm sph}$, the inflowing gas could be quasi-isotropic. If such a region were present within the PBH sphere of influence, the supersonic outflow should have enough ram pressure to overcome that of the inflowing gas, if distances $>r_{\rm acc}$ are to be reached. One can estimate when it is the case assuming that accretion is roughly spherical above $h_{\rm sph}$ from the PBH, that the gas flows inwards at free fall velocity, and then deriving a constraint on $\epsilon$ by equating the outflow and medium ram pressures, $p_{\rm w}=p_{\rm m}$ at $h_{\rm sph}$:
\begin{equation}
2L_{\rm w}/v_{\rm w}\pi h_{\rm sph}^2\tan^2\chi=(\dot{M}_{\rm PBH}/4\pi h_{\rm sph}^2)\sqrt{2GM_{\rm PBH}/h_{\rm sph}}\,
\end{equation}
or
\begin{equation}
L_{\rm w}/c\pi h_{\rm sph}^2\tan^2\chi=(\dot{M}_{\rm PBH}/4\pi h_{\rm sph}^2)\sqrt{2GM_{\rm PBH}/h_{\rm sph}}\,
\end{equation}
for the non-relativistic and the highly relativistic case, respectively; $\chi$ is the outflow half opening angle.
Thus, the outflow will escape if:
\begin{equation}
\begin{aligned}
\epsilon &\gtrsim (v_{\rm w}/c)\chi^2(GM_{\rm PBH}/32c^2h_{\rm sph})^{1/2} \\ & \approx 2\times 10^{-7}(v_{\rm w}/0.1\,c)\chi_{-1}^2(h_{\rm sph}/10^6\,r_{\rm g})^{-1/2} 
\end{aligned}
\end{equation}
or
\begin{equation}
\epsilon\gtrsim \chi^2(GM_{\rm PBH}/8c^2h_{\rm sph})^{1/2}\approx 4\times 10^{-6}\chi_{-1}^2(h_{\rm sph}/10^6\,r_{\rm g})^{-1/2} 
\end{equation}
for the non-relativistic and the highly relativistic case, respectively, where $\tan\chi$ has been approximated as $\chi$ (valid for $\chi\ll 1$). 


\section{Mechanical feedback: analytical estimates}
\label{sec:mechanical_feedback_analytic}

In this section, we characterize analytically the interaction of the outflow with the surrounding medium in its early (Sect.~\ref{subsec:early_interaction}) and long-term (Sects.~\ref{subsec:perpendicular_outflow} and~\ref{subsec:parallel_outflow}) stages of evolution. We discuss the effect of the relative orientation between the outflow and the PBH velocity, corresponding to different values of the angle $\theta$ shown in Fig.~\ref{fig:sketch_early_stage}. We investigate the cases when the outflow and the PBH velocities are orthogonal ($\theta=\pi/2$; Sect.~\ref{subsec:perpendicular_outflow}), and when they are aligned ($\theta=0$; Sect.~\ref{subsec:parallel_outflow}). In the long-term evolution, the value of the half-opening angle $\chi$ is expected to be also important for mechanical feedback; we discuss two qualitatively distinct cases: collimated outflows with $\chi\ll 1$ (see Sects.~\ref{subsec:perpendicular_outflow} and~\ref{subsec:parallel_outflow}) versus non-collimated outflows with $\chi\sim 1$ (see Sect.~\ref{subsec:parallel_outflow}). In this section we consider as reference case a conical outflow with half-opening angle $\chi = 10^{-1}$, velocity of the outflow $v_{\rm w} = 10^9\ {\rm cm\, s^{-1}}$, and luminosity $L_{\rm w} = 10^{30}\ {\rm erg\, s^{-1}}$, corresponding to $\epsilon\approx 10^{-4}$ (see Eq.~\ref{eq:outflow_luminosity}) for our choice of parameters. Nevertheless we note that our predictions will not be very sensitive to the actual parameter values (with the clear exception of $\chi$) as long as $L_{\rm w}$ is large enough to allow the outflow front to escape the PBH sphere of influence.


\subsection{Early interaction} 
\label{subsec:early_interaction}

\begin{figure}        
\includegraphics[width=\columnwidth]{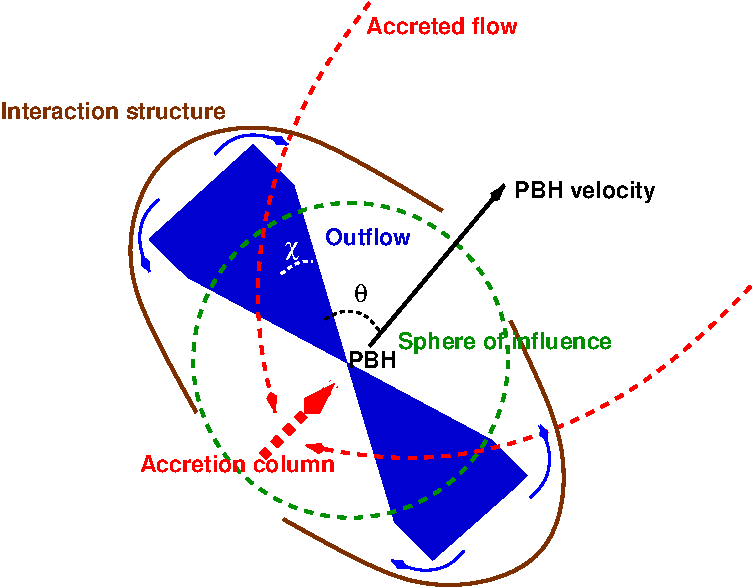}
\caption{Sketch of the early stage of the outflow-medium interaction, including the formation of a diluted and hot gas bubble that ultimately engulfs the PBH sphere of influence. Sketch not to scale.}
\label{fig:sketch_early_stage}
\end{figure}

After leaving the region close to the compact object, the front of the supersonic outflow is eventually slowed down by the ram pressure of the gas surrounding the PBH. Since the medium sound speed is significantly smaller than $v_{\rm w}$, the medium ram pressure in the outflow reference frame is $p_{\rm m}\approx \rho_{\rm m} v^2_{\rm w}$. The ram pressure of the outflow front in the medium frame is:
\begin{equation}
p_{\rm w} \approx \frac{2L_{\rm w}}{S v_{\rm w}} = \frac{2L_{\rm w}}{\pi  h^2 \tan^2\chi v_{\rm w}},
\end{equation}
in the non-relativistic case, where $S = \pi h^2 \tan^2\chi$ is the outflow front surface at the distance $h$ from the PBH. For typical values of the parameters we have that the outflow and medium pressures become comparable at a height:
\begin{equation}
h \approx 2\times 10^{14} L_{\rm w,30}^{1/2} n_{\rm m,0}^{-1/2} \chi^{-1}_{-1} v_{\rm w,9}^{-3/2}\ {\rm cm},
\label{eq:height_early_interaction}
\end{equation}
where $\tan\chi \sim \chi$. If the outflow velocity were highly relativistic, then $p_{\rm m}\approx\rho_{\rm m}\gamma^2_{\rm w} v^2_{\rm w}$ and $p_{\rm w}\approx L_{\rm w}/(S v_{\rm w})$, turning $h$ into $h^{\rm rel}=h/(\gamma_{\rm w}\sqrt{2})$. 

Approximately at the distance $h$ from the PBH, where pressure equilibrium
is reached, the front of the supersonic outflow significantly slows down and the
material reaching the front gets shocked. Then, this shocked material is
directed sideways and then backwards, and inflates a bubble of hot adiabatic gas surrounded by a shell of shocked medium.
The outflow is initially freely expanding, but at some point the bubble pressure balances the lateral
pressure of the outflow, point at which the latter acquires a cylindrical geometry. This leads to the
self-similar evolution of the bubble, in which the bubble dynamics depends on its energy content and the accumulated medium mass \citep[see, e.g.,][in an extragalactic context]{ale06}. The bubble evolution is thus not very
sensitive to how collimated or relativistic the initial outflow was. This process is illustrated in Fig.~\ref{fig:sketch_early_stage}.

The self-similar bubble expansion can be approximated as adiabatic and symmetric, fed by a steady source of energy; thus, its evolution can be approximately characterized by the Sedov-Taylor equation in the continuous regime (see, e.g., \citealt{bla76} and references therein; see also \citealt{kai97} and \citealt{ale06} in the context of extragalactic jets), according to which the bubble radius $R_{\rm b}$ evolves in an approximate self-similar fashion as:
\begin{equation}
\begin{aligned}
R_{\rm b} (t)\sim \left(\frac{L_{\rm w}}{\rho_{\rm m}}\right)^{1/5}\,t^{3/5}\approx 3\times 10^{14}\,L_{\rm w,30}^{1/5}\,t_7^{3/5}\,n_{\rm m,0}^{-1/5}\ {\rm cm},
\end{aligned}
\label{eq:bubble_radius}
\end{equation}
with forward velocity:
\begin{equation}
v_{\rm b} (t)=\frac{{\rm d}R_{\rm b}(t)}{{\rm d}t}\sim 2\times 10^7\,R_{\rm b,14.5}\,t_7^{-1}\ {\rm cm~s}^{-1}.
\label{eq:bubble_velocity}
\end{equation}
Thus, for some time the bubble is expanding forward significantly faster than the PBH moves, for $v_{\rm PBH}=3 \times 10^6\ {\rm cm\, s^{-1}}$. Here, for simplicity, we assume that the bubble is approximately spherical, although the lateral expansion can be a factor of a few slower \citep[see, e.g.,][]{kai97}, which should be accounted for in a more detailed treatment of the problem.

If the pressure equilibrium distance is larger than the effective accretion radius, that is, $h>r_{\rm acc}$, the bubble will start to form after a time $t_{\rm h}\sim h/v_{\rm w}$, and once the outflow front has already left the PBH sphere of influence. In this case, the time required by the front to cross the sphere of influence $t_{\rm cr}$ roughly corresponds to the time needed to reach a distance $r_{\rm acc}$ from the PBH: $t_{\rm cr}\sim 10^{6}\ r_{\rm acc,15}/v_{\rm w,9}\ {\rm s}$ ($<t_{\rm h}$). Viceversa, if $h<r_{\rm acc}$, the crossing time will be $t_{\rm cr} \sim r_{\rm acc}/v_{\rm b}(t_{\rm cr}) \approx 10^7\ r^{5/3}_{\rm acc, 15} n^{1/3}_{\rm m,0} L^{-1/3}_{\rm w,30} \ {\rm s}\gg r_{\rm acc}/v_{\rm w}$. In either case, after a time $\gtrsim \max[t_{\rm h},t_{\rm cr}]$, the bubble filled with hot gas will engulf the PBH sphere of influence and accretion is expected to be affected. Only if $v_{\rm b}$ becomes $\lesssim v_{\rm PBH}$ before the bubble reaches out of the PBH sphere of influence (i.e., an extreme case plus the condition $h<r_{\rm acc}$), bubble growth will be overcome by gravity, and the outflow will not be able to prevent accretion. 


\subsection{Quasi-perpendicular outflow} 
\label{subsec:perpendicular_outflow}

\begin{figure}        
\includegraphics[width=\columnwidth]{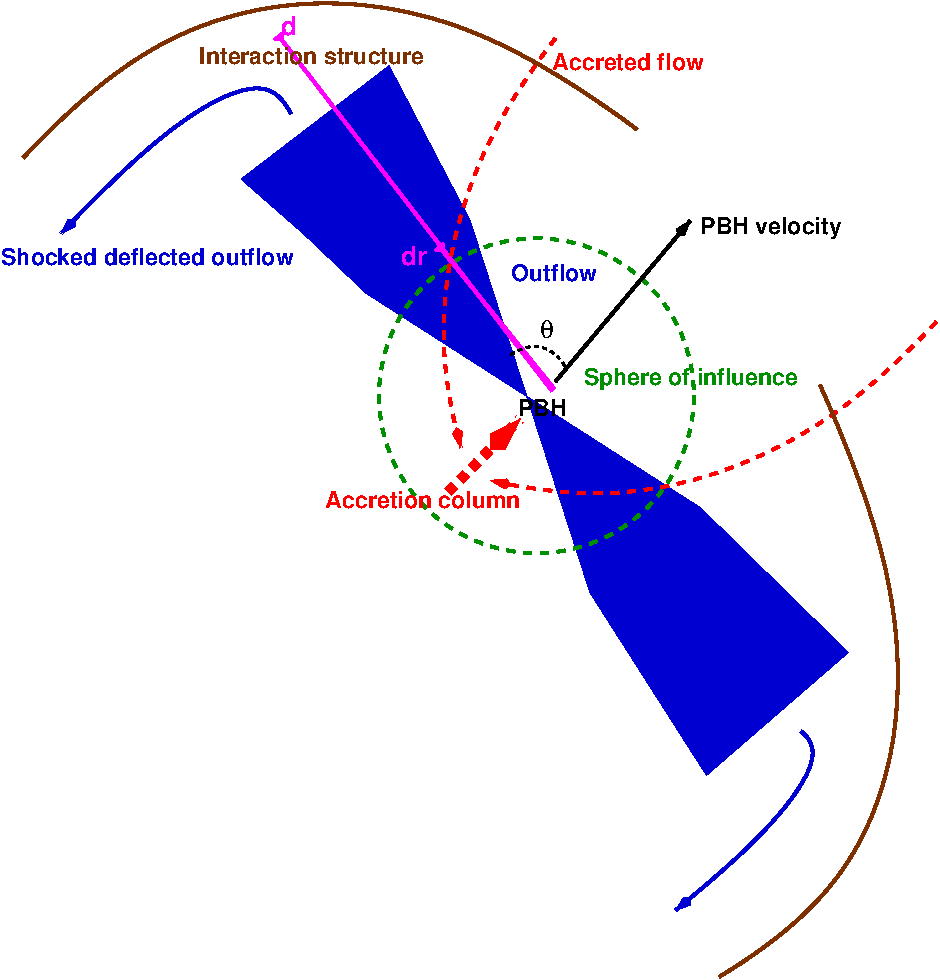}
\caption{Long-term outflow-medium interaction in the quasi-perpendicular case. Sketch not to scale.}
\label{fig:sketch_longterm_perpendicular}
\end{figure}

Once outside the PBH sphere of influence, the bubble slows down until its expansion velocity and the PBH velocity become comparable, that is, $v_{\rm b}\sim v_{\rm PBH}$. At that moment and for $\theta\sim 1$, the bubble is already far from symmetric because the medium ram pressure is mostly acting on one side of the bubble. The typical timescale of the previous axisymmetric expansion $t_{\rm ax}$ can be derived from Eq.~\eqref{eq:bubble_velocity}:
\begin{equation}
t_{\rm ax} \sim \sqrt{\left(\frac{3}{5}\right)^{5}\frac{L_{\rm w}}{\rho_{\rm m}v^5_{\rm PBH}}} \\
\approx 1.2 \times 10^{10} L_{\rm w,30}^{1/2}\,n_{\rm m,0}^{-1/2}\,v_{\rm PBH,6.5}^{-5/2}\ {\rm s},
\end{equation}
and the bubble size at time $t_{\rm ax}$ is:
\begin{equation}
\begin{aligned}
R_{\rm b} (t_{\rm ax}) & \sim \frac{5}{3} v_{\rm PBH} t_{\rm ax} = \sqrt{\left(\frac{3}{5}\right)^{3} \frac{L_{\rm w}}{\rho_{\rm m}v^3_{\rm PBH}}}\\
& \approx 5 \times 10^{16} v_{\rm PBH,6.5} t_{\rm ax,10} \ {\rm cm}.
\label{qprb}
\end{aligned}
\end{equation}

Therefore, for the system under consideration, we have that after $t\sim t_{\rm ax}\approx 400$~yr the bubble will reach a typical size of $\sim 60\,r_{\rm acc}$ (see Eqs.~\ref{eq:accretion_radius} and \ref{qprb}). At that point, the bubble is already getting compressed in the direction of the PBH motion, and tends to expand in the opposite direction, as shown in Fig.~\ref{fig:sketch_longterm_perpendicular} (see also, e.g., \citealt{yoo11} for simulations of jets from a BH microquasar moving in the interstellar medium, and \citealt{li19} for a simulation of a case more similar to ours). 

The strong bubble asymmetry makes the two constituents of the bipolar outflow deflect away from the PBH motion direction. The typical scale $d$ at which the outflow deflection is significant can be derived from equating the medium ram pressure in the PBH reference frame, $p_{\rm m}\approx \rho_{\rm m}v^2_{\rm PBH}$, with the outflow ram pressure, $p_{\rm w}\approx 2L_{\rm w} / v_{\rm w} \pi d^2 \chi^2$ in the non-relativistic case:
\begin{equation}
d\approx 7\times 10^{16} L_{\rm w,30}^{1/2}  n_{\rm m,0}^{-1/2} v_{\rm w,9}^{-1/2} v_{\rm PBH,6.5}^{-1} \chi^{-1}_{-1} \ {\rm cm},
\label{eq:height_longterm_evolution}
\end{equation}
where we assume that $\chi\ll 1$. In the highly relativistic case, for which $p_{\rm w}\approx L_{\rm w} / c \pi (d^{\rm rel})^2 \chi^2$, this distance turns into $d^{\rm rel}=d/\sqrt{2}$. The deflection distance $d$ turns out to be much larger than the PBH sphere of influence; for the parameters considered in this work: $d\sim 70\,r_{\rm acc}$. This scale also characterizes the typical size of the section of the interaction structure perpendicular to the PBH motion. 

For $\theta=\pi/2$ (and $\chi\ll 1$), free sidewards expansion of the outflow propagating within the asymmetric bubble is stopped on the side on which the medium ram pressure acts, on a distance $d_{\rm r}<d$ from the PBH. Assuming for simplicity that the outflow is very cold, outflow asymmetric reconfinement occurs at the distance where the medium and outflow lateral ram pressures are equal, $\rho_{\rm m}v_{\rm PBH}^2=2L_{\rm w}/\pi d_{\rm r}^2v_{\rm w}$ in the non-relativistic case:
\begin{equation}
d_{\rm r}\sim \chi\,d\approx 
7\times 10^{15}\,L_{\rm w,30}^{1/2}\,n_{\rm m,0}^{-1/2}\,v_{\rm w,9}^{-1/2}\,v_{\rm PBH,6.5}^{-1}\,{\rm cm}.
\end{equation}
In the highly relativistic case, where $\rho_{\rm m}v_{\rm PBH}^2=L_{\rm w}/\gamma_{\rm w}^2c\pi (d^{\rm rel}_{\rm r})^2$, this distance becomes $d^{\rm rel}_{\rm r}=d_{\rm r}/(\gamma_{\rm w}\sqrt{2})$. Within $d_{\rm r}$, the outflow is expected to be ballistic, at $\sim d_{\rm r}$ it starts to deflect, and at $d$, it has significantly deviated. The tail described above is in fact made of shocked deflected outflow heavily mixed with shocked medium, as the outflow-medium contact discontinuity is prone to hydrodynamical instability growth (see, e.g., Sect.~\ref{sec:mechanical_feedback_simulations}). 

As shown by \cite{li19}, for $\theta=\pi/2$ the surrounding medium gas can penetrate into the sphere of influence and reach the PBH on the directions perpendicular to the outflow, because the bubble pressure is the lowest there. Thus, accretion may only be moderately affected by mechanical feedback. Nevertheless we note that for $\chi\ll 1$ and $\theta<\pi/2$ (but still $\sim 1$), the outflow can suffer a stronger, oblique shock that can widen the outflow front. This situation might turn into a case where we effectively have $\chi\sim\theta$, which is closer to the quasi-parallel case reported in Sect.~\ref{subsec:parallel_outflow}.

So far we have assumed that the evolution of the outflow-medium interaction structure was adiabatic. However, according to the cooling rates for low metallicities in \cite{sut93} and the timescale of the problem ($\gtrsim t_{\rm ax}$) close the recombination epoch the medium was dense enough ($n_{\rm m}(z_{\rm rec} \approx 1100) \approx 400\ {\rm cm^{-3}}$) to make the bubble-driven shock in the medium radiative. This affects the dynamics of the bubble; in particular, it enhances the growth of instabilities at the outflow-medium contact discontinuity. A fully disrupted contact discontinuity would make an analytical treatment of the studied scenario more speculative. At this stage, we leave for future work a more detailed analysis of the radiative bubble case.


\subsection{Quasi-parallel outflow} 
\label{subsec:parallel_outflow}

\begin{figure}     
\includegraphics[width=\columnwidth]{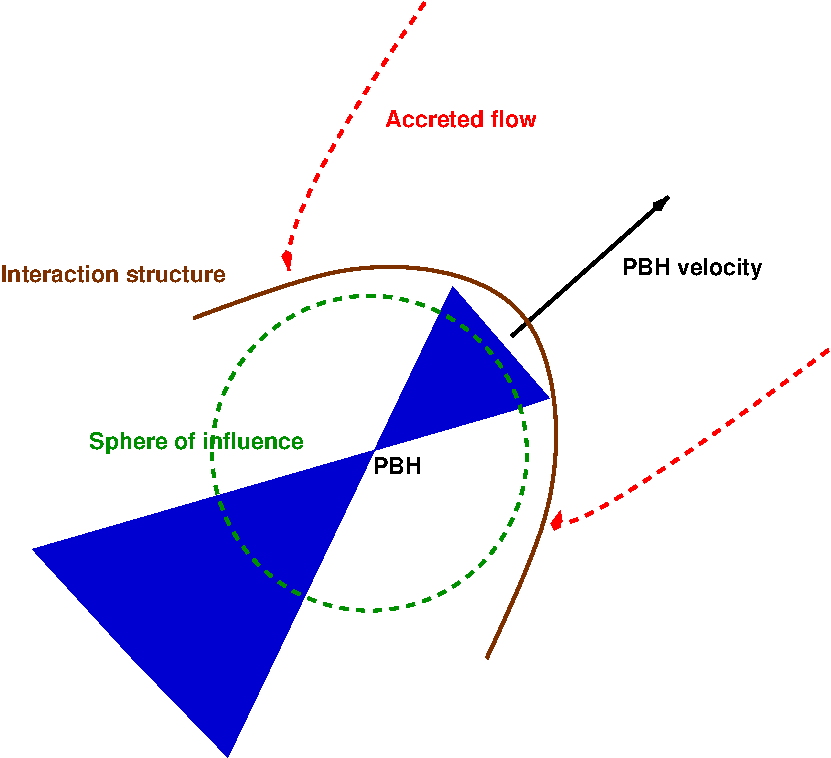}
\caption{Long-term outflow-medium interaction in the quasi-parallel case. Sketch not to scale.}
\label{fig:sketch_longterm_parallel}
\end{figure}

When the outflow is roughly aligned with the direction of motion (see Fig.~\ref{fig:sketch_longterm_parallel}), accretion onto the PBH sphere of influence is expected to be almost halted once the hot and diluted bubble gas fully occupies that region, keeping out the external gas (as shown in Sect.~\ref{sec:mechanical_feedback_simulations}, and also in \citealt{li19}). As already mentioned, this could occur also for non-negligible values of $\theta$ if the outflow is non-collimated, that is, $\chi\sim 1$. Under these circumstances, significant outflow production can last only until the remaining material within the sphere of influence has been accreted. 

The typical timescale of accretion of the remaining medium gas is of order of the free-fall timescale at $r_{\rm acc}$:
\begin{equation}
t_{\rm acc} \sim \frac{r_{\rm acc}}{v_{\rm PBH}} \approx 3\times 10^8 r_{\rm acc,15} v_{\rm PBH,6.5}^{-1}\ {\rm s}.
\label{eq:accretion_timescale}
\end{equation}
After a time $\sim t_{\rm cr}+t_{\rm acc}$ from the formation of the outflow, only diluted hot matter should be present around the PBH and accretion should be very weak. The bubble size and expansion speed at this stage are given by Eqs.~\eqref{eq:bubble_radius} and~\eqref{eq:bubble_velocity}, and they read as:
\begin{equation}
R_{\rm b} (t_{\rm cr}+t_{\rm acc})\approx 1.4\times 10^{16} L_{\rm w,30}^{1/5} t_9^{3/5} n_{\rm m,0}^{-1/5}\ {\rm cm}
\end{equation}
and
\begin{equation}
v_{\rm b} (t_{\rm cr}+t_{\rm acc})\approx 2\times 10^7 R_{\rm b,16} t_9^{-1}\ {\rm cm~s^{-1}},
\end{equation}
respectively. Therefore, even after having expanded up to $\sim 20\,r_{\rm acc}$, the bubble expansion speed is still significantly faster than the PBH, for $v_{\rm PBH}=3 \times 10^6\ {\rm cm\, s^{-1}}$. 

When both accretion and ejection stop, energy injection into the bubble stops as well. At that point, the outflow has injected an energy of:
\begin{equation}
E_{\rm inj} \sim L_{\rm w} \times (t_{\rm cr}+t_{\rm acc}) \approx 10^{39} L_{\rm w,30} t_{\rm 9}\ {\rm erg}.
\end{equation}
Once the energy injection stops, for $t\gg (t_{\rm cr}+t_{\rm acc})$ the bubble will expand adiabatically with size and velocity approximated as \citep[see, e.g.,][for the adiabatic and impulsive regime]{bla76}:
\begin{equation}
R_{\rm b}(t) \sim \left(\frac{E_{\rm inj}}{\rho_{\rm m}}\right)^{1/5} t^{2/5}
\end{equation}
and
\begin{equation}
v_{\rm b}(t) \sim \frac{2}{5}\left(\frac{E_{\rm inj}}{\rho_{\rm m}}\right)^{1/5} t^{-3/5},
\end{equation}
respectively.

The time needed for the bubble velocity $v_{\rm b}$ to reach the PBH velocity $v_{\rm PBH}$ reads as:
\begin{equation}
\begin{aligned}
t_{\rm end} &\sim \left(\frac{2^5}{5^5}\frac{E_{\rm inj}}{\rho_{\rm m}v_{\rm PBH}^{5}}\right)^{1/3} \\
&\approx 3\times 10^{9} E_{\rm inj,39}^{1/3} n_{\rm m,0}^{-1/3} v_{\rm PBH,6.5}^{-5/3}\ {\rm s},
\end{aligned}
\end{equation}
and the bubble final size after a time $t_{\rm cr}+t_{\rm acc}+t_{\rm end}\sim t_{\rm end}$ is:
\begin{equation}
R_{\rm b}(t_{\rm end}) \sim \frac{5}{2} v_{\rm PBH} t_{\rm end} \approx 3 \times 10^{16} v_{\rm PBH,6.5} t_{\rm end,9.5}\ {\rm cm},
\label{eq:bubble_final_radius} 
\end{equation}
which is somewhat smaller than the value obtained in Eq.~\eqref{eq:height_longterm_evolution} for the quasi-perpendicular case, but still much larger than the PBH sphere of influence. 

Once $v_{\rm b}$ gets $\lesssim v_{\rm PBH}$, the ram pressure of the medium becomes dominant over that of the shocked outflow, and the material starts re-approaching the PBH as the latter moves, reaching it in an accretion recovery time: 
\begin{equation}
t_{\rm rec} \sim R_{\rm end}/v_{\rm PBH} \approx 10^{10} R_{\rm end,16.5} v_{\rm PBH,6.5}\ {\rm s}.
\end{equation}
As $(t_{\rm rec}+t_{\rm end})\gg (t_{\rm cr}+t_{\rm acc})$, the stage in which the PBH neither accretes nor ejects largely dominates the duration of the whole on-off accretion cycle, and so the time-average accretion rate should be much lower than the Bondi-Hoyle-Lyttleton one. For the typical values derived here, we find an average reduction in $\dot{M}_{\rm PBH}$ by a factor $(t_{\rm cr}+t_{\rm acc})/(t_{\rm rec}+t_{\rm end})\sim 0.1$ (see Sect.~\ref{sec:mechanical_feedback_simulations}), but lower and higher values are also possible given the large plausible parameter space.

The maximum, temporary section radius of the outflow-medium interaction structure in the quasi-parallel case can be identified with the bubble radius $R_{\rm b}(t_{\rm end})$. If otherwise accretion and outflow production were constant, the structure section radius would be also constant and characterized by Eq.~\eqref{eq:height_longterm_evolution}.


\section{Mechanical feedback: numerical calculations}
\label{sec:mechanical_feedback_simulations}

In Sect.~\ref{subsec:perpendicular_outflow} we have seen that a perpendicular, well-collimated outflow may leave the PBH sphere of influence without strongly affecting the gas accretion from directions away from the outflow, allowing the gas to reach the PBH. On the other hand, even in the perpendicular case, accretion could be reduced if high-pressure, shocked outflow material were more homogeneously distributed inside the PBH sphere of influence, more effectively preventing the external medium from entering that region. As noted, this could be the case for more oblique and (or) broad outflows, but even collimated ($\chi\ll 1$) perpendicular outflows may in principle divert enough momentum sideways due to instability growth closer to the PBH, leading to a situation closer to that discussed in Sect.~\ref{subsec:parallel_outflow}. Unfortunately, finding the value range of $\theta$ and $\chi$ for which accretion is inhibited by mechanical feedback requires a number of expensive 3-dimensional (3D) hydrodynamical simulations, as they should include spatial scales both $\ll r_{\rm acc}$ and $\gtrsim d$; for typical parameter values this spans several orders of magnitude. Adaptive mesh refinement can be useful to reduce the computational time, but we note that highly turbulent flows, as expected in the studied scenario, are sometimes challenging for this technique \citep[see, e.g.,][]{bos15}.

The computing time needed for a simulation is proportional to the number of computing cells of the grid, times the number of simulation time steps: $T_{\rm sim} \propto N_{\rm cells} \times N_{\rm steps}$, where $N_{\rm cells} \propto (d/r_{\rm acc})^D \propto (v_{\rm w}/v_{\rm PBH})^{D/2}$, with $D$ being the dimensionality, and $N_{\rm steps} \propto t_{\rm phys}/\Delta t_{\rm phys}$, where $t_{\rm phys}\propto d/v_{\rm PBH}$ is the total simulated time, and $\Delta t_{\rm sim} \propto d/v_{\rm w}$ the simulated time step. This yields $T_{\rm sim} \propto (v_{\rm w}/v_{\rm PBH})^{D/2+1}$. To make the problem more tractable, one can choose parameter values to reduce $T_{\rm sim}$, although one still requires that $d/r_{\rm acc}\gg 1$ to capture at least qualitatively the main features of the system. This additional constraint makes a 3D simulation quite demanding. On the other hand, a 2D simulation, which is realistic in the parallel outflow case because of axisymmetry, becomes more affordable. 

For this work, as a test of the case in which accretion is strongly reduced due to mechanical
feedback, we carried out axisymmetric hydrodynamical simulations for $\theta=0$. We adopted
values for $v_{\rm PBH}$ and $v_{\rm w}$ that allow us to probe a moderately large range of
spatial scales $d/r_{\rm acc}$, and that reduce significantly $N_{\rm steps}$. An intermediate
value of the outflow half-opening angle was obtained by adopting a Mach number of $M_{\rm ch}=2.5$,
which yields shortly after injection, already in the supersonic regime, a $\chi\approx 1/M_{\rm ch}=0.4$; but we note that trials with $\chi=0.1$
(not shown here) yielded similar results. 


\subsection{Properties of the simulation}

We performed axisymmetric hydrodynamical simulations of the interaction between a collimated outflow with $v_{\rm w}=10^9\ {\rm cm\, s^{-1}}$ and initial $M_{\rm ch}=2.5$, and a moving medium with $v_{\rm PBH}=5\times 10^7$~cm~s$^{-1}$ and Mach number of 3. The medium density was fixed to $\rho_{\rm m}=1.7\times 10^{-24}\ {\rm cm^{-3}}$ (i.e., $n_{\rm m}\approx 1\ {\rm cm^{-3}}$), and the power of each constituent of the bipolar outflow to $L_{\rm w}=0.025\dot{M}_{\rm PBH}v_{\rm w}^2$, corresponding to $\epsilon\approx 10^{-4}$, which is the reference value used for our analytical estimates above. The grid size was chosen such that the interaction structure remained in the computational grid for the duration of the simulation. We note that exploratory trials done with a higher $\epsilon$-value suggested that results should not change significantly. 

The cell size of the simulation was fixed to $10^{11}\ {\rm cm}$. The size of the grid in the
radial $\hat{r}$- and the vertical $\hat{z}$-directions are 250 cells ($2.5\times 10^{13}\
{\rm cm}$) and 500 cells ($5\times 10^{13}\ {\rm cm}$), respectively. The source of the
gravitational potential is assumed to be point-like, with mass $M_{\rm PBH} = 30 $~M$_\odot$,
but the grid region where matter is accreted and leaves the grid (i.e., the accretor in the context of the numerical simulation) is
modeled as a spherical region of low density and pressure, with a radius of 10 cells and
centred at the origin. The outflow is injected in the $\pm\hat{z}$-directions through a
cylindrical inlet with a height and radius of 20 and 5 cells, respectively, and centred also
at the origin, occupying thus a fraction of the accretion sphere. Any material crossing the
accretor spherical surface but at the outflow inlet boundaries immediately disappears from
the grid. Outside the outflow inlet and the accretor region, reflection conditions are imposed
at the vertical axis ($r=0$: left grid boundary), outflow conditions\footnote{The flow properties at the boundary cells are set equal to those at the cells next to the boundary, and inside the computational domain. The flow leaves or enters the grid depending on the sign of the velocity components normal to the boundary planes.} at the upper
and right grid boundaries, and inflow conditions (i.e., the medium) at the bottom grid
boundary. The motion of the medium in the PBH reference frame is in the $+\hat{z}$-direction.
The accretion rate is computed on the cells right outside the boundary of the accretor
(i.e., excluding the outflow inlet). In this specific set-up the sphere of influence has
approximately a radius of 30 cells ($3\times 10^{12}\ {\rm cm}$). Trials carried out with
better resolution by a factor $\approx 1.9$, and a slightly smaller accretor radius, $\approx
26$\% instead of $30$\%, yielded very similar results to those presented here.

The code used to solve the hydrodynamics equations is the same as the one used in \cite{del17}: third order in space~\citep{mig05}; second order in time; and using the Marquina flux formula \citep[in the non-relativistic case,][]{don96}. Both the outflow and medium fluids were simplified as non-relativistic mono-atomic gases with adiabatic index $5/3$. We neglect the effects of ionization and gas abundances, which are not relevant on the simulated scales ($\gg r_{\rm g}$), on which gravity and ram pressure dominate the dynamics of an adiabatic flow. Any magnetic field should be affected by the flow evolution, but here we are assuming that the role of the magnetic field is purely subsidiary to the collisionless gas hydrodynamical approximation, and so of no dynamical importance. The accretor gravitational force was appropriately included in the hydrodynamical equations as a source term \citep[see, e.g.,][]{tor09}.


\subsection{Results}
\label{subsec:numerical_results}

\begin{figure}        
\includegraphics[width=\columnwidth]{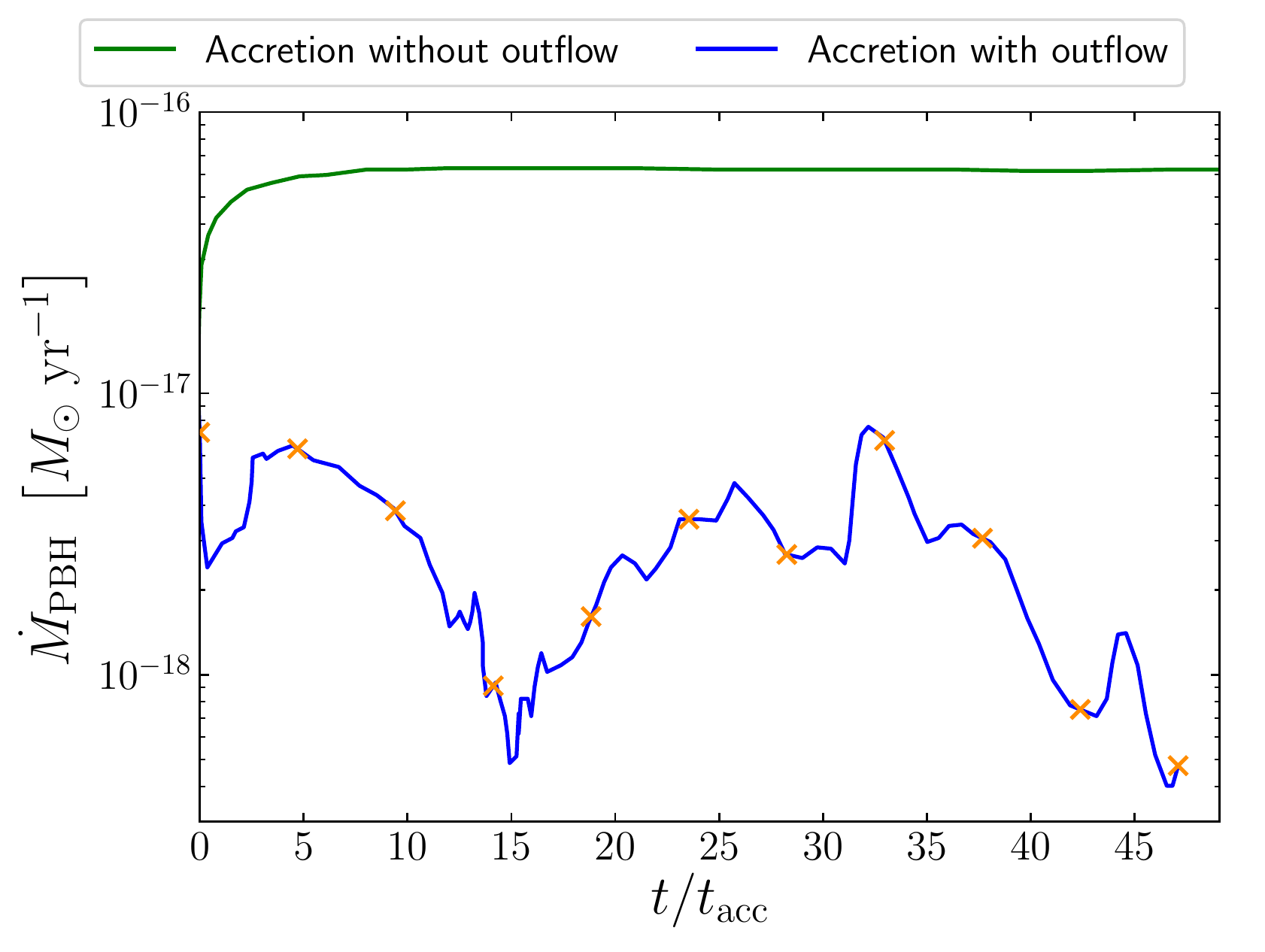}
\caption{Evolution of the accretion rate with (blue curve) and without outflow (green curve). Time is measured in units of the accretion dynamical timescale~$t_{\rm acc} = 6\times 10^4\ {\rm s}$. Orange crosses are located at the times of the eleven density maps with outflow of Fig.~\ref{fig:density_maps}.}
\label{fig:accretion_rate}
\end{figure}

We show in Fig.~\ref{fig:accretion_rate} the accretion rate of two cases: when the mechanical
feedback mechanism due to the outflow is present and when it is not. When there is no
feedback, the long-term accretion rate converges to a value within a factor of 2 from
$\dot{M}_{\rm PBH}\approx 3\times 10^{-17}$~M$_\odot$~yr$^{-1}$ predicted by
Eq.~\eqref{eq:bondi_hoyle_accretion_rate}. On the other hand, when outflows are launched, our
simulations indicate that matter reaches the PBH only indirectly, through filamentary
structures related to hydrodynamical instabilities in the contact discontinuity between the
outflow and the medium. Thus, accretion is on average one order of magnitude lower
($\dot{M}_{\rm PBH}\approx 3\times 10^{-18}$~M$_\odot$~yr$^{-1}$) than in the case without
outflow, but presents strong fluctuations with respect to that average value. It is worth
indicating the presence of filamentary material entering the grid from the top for $t\gtrsim
15\,t_{\rm acc}$. This backflow is a numerical artifact associated to the boundary
conditions and the presence of subsonic material in the upper boundary. The subsonic material in the boundary can induce the flow to bounce back, sending waves into the grid. However, we have
checked with additional simulation trials that imposing a positive velocity at the boundary (i.e., the flow is forced to leave the grid), of modulus $5\times 10^7$~cm~s$^{-1}$,
quantitatively yields very similar results. The simulation was run until it presented
repetitive patterns of the flow, a sort of quasi-steady state. Our results are quite similar
to those obtained by \cite{li19} for $\theta=0$.

\begin{figure*}        
       \centering       
       \includegraphics[width=4.4cm]{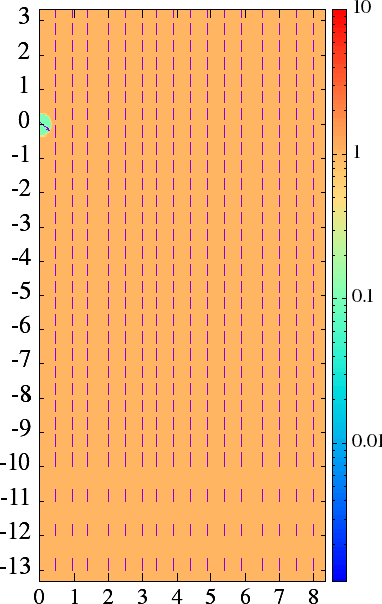}
       \includegraphics[width=4.4cm]{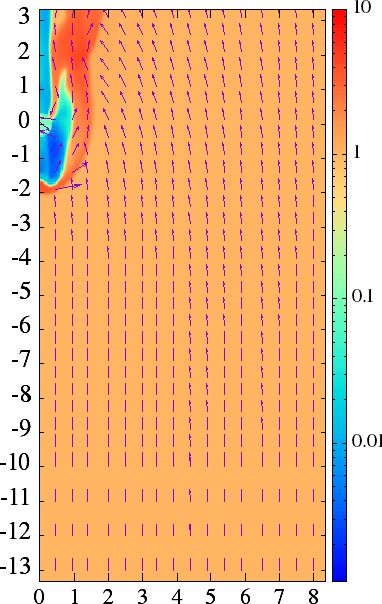}
       \includegraphics[width=4.4cm]{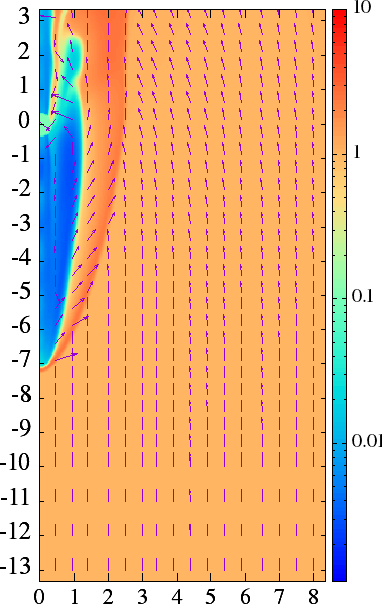}
       \includegraphics[width=4.4cm]{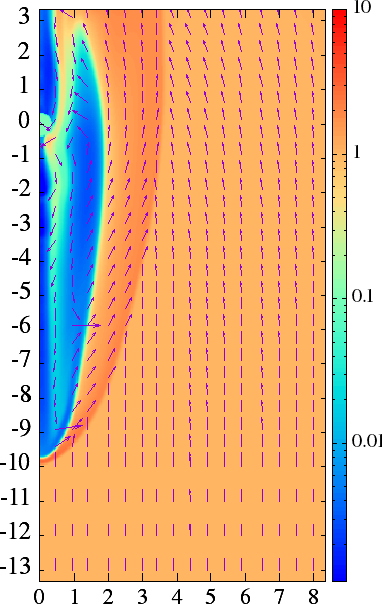}\\      
       \includegraphics[width=4.4cm]{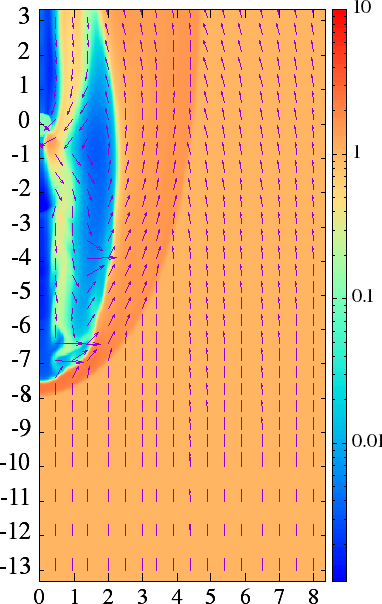}
       \includegraphics[width=4.4cm]{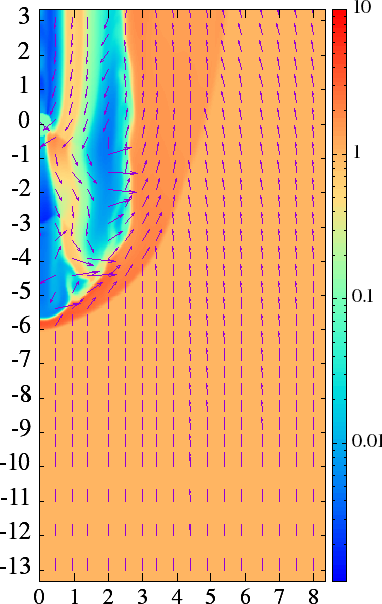}      
       \includegraphics[width=4.4cm]{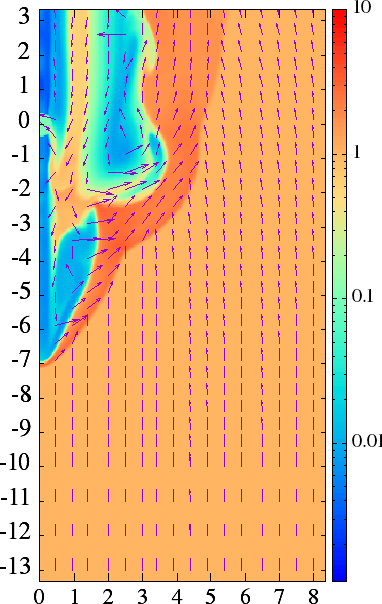}      
       \includegraphics[width=4.4cm]{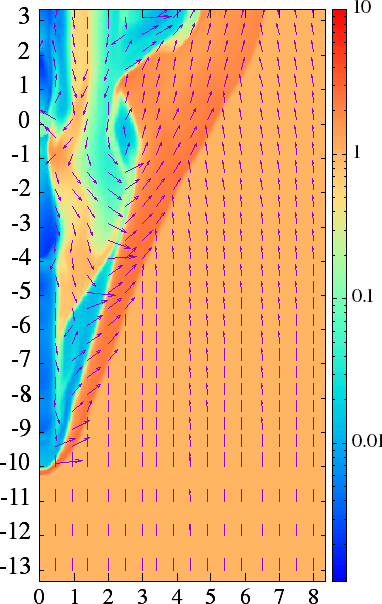}\\
       \includegraphics[width=4.4cm]{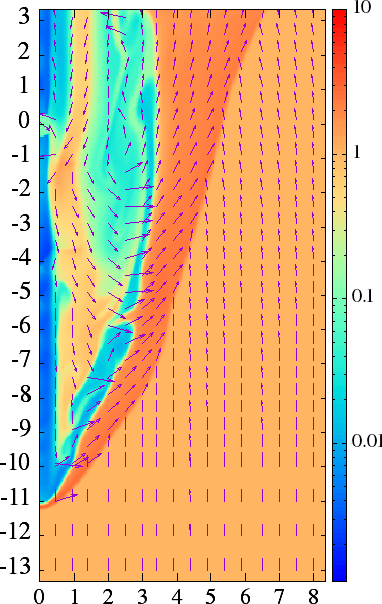}     
       \includegraphics[width=4.4cm]{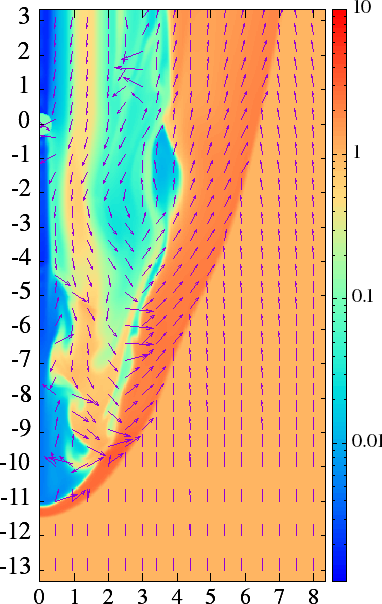}      
       \includegraphics[width=4.4cm]{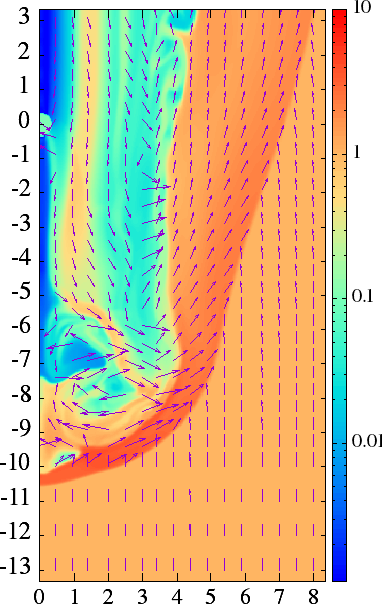}
       \includegraphics[width=4.4cm]{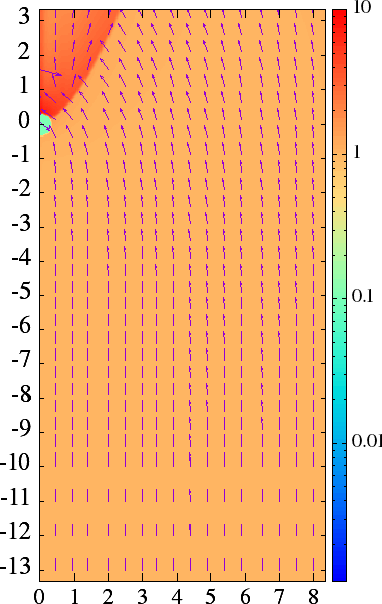}
       \caption{Color density maps of the outflow-medium interaction structure after (from left to right, and top to bottom) $t\approx 0$, 5, 10; 15, 20, 25; 30, 35, 40; 45, $50\,t_{\rm acc}$. The bottom right density map corresponds to the case without outflow after $t\approx 50\,t_{\rm acc}$. The arrows illustrate the accreted gas trajectories. The injector-accretor region is located at $(0,0)$. The axis scales are normalized to $r_{\rm acc}$, and the color scale is normalized to the medium density. The medium motion in the PBH reference frame is upwards.}
       \label{fig:density_maps}
\end{figure*}

We also present in Fig.~\ref{fig:density_maps} the density maps showing the evolution of the outflow structure for approximately 50 times the accretion dynamical timescale $t_{\rm acc}$ ($t_{\rm acc}\approx 6\times 10^4\ {\rm s}$, according to Eq.~\ref{eq:accretion_timescale}). The time of each snapshot in Fig.~\ref{fig:density_maps} is marked with a cross at its corresponding location in the accretion rate curve of Fig.~\ref{fig:accretion_rate}. These density maps can be compared to the no-outflow simulation, run for a very similar amount of time, and shown in the bottom right panel of Fig.~\ref{fig:density_maps}. Both Figs.~\ref{fig:accretion_rate} and~\ref{fig:density_maps} are illustrative of the strong non-linear nature of the outflow-medium interaction structure, and reminiscent of the intermittent nature of accretion predicted in Sect.~\ref{sec:mechanical_feedback_analytic}. 

\subsection{A toy-model for $\theta=\pi/2$}\label{toy}

Despite the difficulty of studying the quasi-perpendicular case numerically, we still made a
toy simulation in 2D of the case $\theta\sim 1$ by introducing an additional gravitational
potential with a velocity dispersion $\sigma$, and with the medium initially at rest \citep[as
under the presence of a dark matter halo; see, e.g.,][for a calculation in the supermassive black-hole case]{zei19}.
We ran test simulations of that scenario in the context of a PBH of 30~M$_\odot$ and
$\sigma=30$~km~s$^{-1}$ and of a BH of $10^8$~M$_\odot$ and $\sigma=300$~km~s$^{-1}$,
and obtained similar results in both cases to those of \cite{zei19} for the second case, that is,
the accretion rate tended to a value a factor of a few smaller than the Bondi value without
additional gravitational potential and medium inward (or sound) speed $\sigma$ \footnote{We note that our simulations had a resolution twice
lower, a few times larger accretor relatively to $r_{\rm acc}$, and a grid few times
smaller than those in \cite{zei19}, and were done with a different code, but the qualitative
behavior was the same.}. This result cannot be overstressed as it is not 3D, but shows the
robustness of our numerical scheme. Moreover, the derived reduction in the accretion rate is
similar to that found by \cite{li19} when computing a case in actual 3D similar to the one
discussed in Sect.~\ref{subsec:perpendicular_outflow}. This similarity is likely related to
the fact that in both cases, 3D supersonic accretion with outflow and $\theta=\pi/2$, and 2D
subsonic accretion with outflow and adding a dark-matter-like potential, there is a characteristic medium velocity ($v_{\rm pbh}$ and $\sigma$, respectively) leading to reduced but steady
accretion on the directions away from the outflow.  

\section{Discussion and conclusions}
\label{sec:conclusions}

The effect of mechanical feedback is expected to reduce accretion onto a moving massive object, but
quantifying this effect is difficult. In this work, we show that the angle between the
outflow direction and the PBH motion, $\theta$, is likely to determine how far below the
Bondi rate the PBH accretion is. A numerical study of a similar but more general scenario by \cite{li19} reaches results in line with those presented here in the context of PBH. The half-opening angle $\chi$ is also expected to play an important role, as broader outflows are expected to have larger solid angles for which accretion is halted. As our analytical estimates and simulations show, an outflow do not need to be strong in order to reduce accretion; it is enough that the outflow effectively heats and dilutes all the gas in the PBH surroundings on scales $\gtrsim r_{\rm acc}$.

Assessing the level of accretion as a function of $\theta$ and $\chi$ is challenging due to the
complexity of the outflow-medium interaction, and its intrinsic 3D nature in general. Nevertheless, we
can already predict that accretion is expected to be significantly smaller than Bondi-Hoyle-Lyttleton
accretion, with the presence of strong fluctuations, at least for values of $\theta\lesssim \chi$. For
$\theta\gg\chi$, analytical and numerical calculations seem to indicate that the reduction may still be
non-negligible, but with accretion proceeding more steadily. Nevertheless, the larger the ratio $v_{\rm
w}/v_{\rm PBH}$, the larger the region affected by the outflow, in which case there is more room for
some redirection of outflow momentum and energy, caused, for instance, by instability growth on the
outflow walls. This might lead to the formation of a more effective barrier for accretion on scales
similar to $r_{\rm acc}$, even for $\theta\approx \pi/2$ and $\chi\ll 1$. In addition, $\theta$ may
change with time, as explained at the end of Sect.~\ref{accre}, which could effectively lead to a
time-averaged $\chi\sim 1$. For all this, knowing the effect of the time-averaged values of $\theta$
and $\chi$ of a PBH, and their statistical distribution in the PBH population(s), is crucial to
estimate the effects of their accretion and related emission. 

It is clear that the inclusion of mechanical feedback on a PBH population has a
direct impact on PBH abundance constraints, as they depend on the energy injected by the PBH population
into the primordial medium. This can be exemplified by the following. In the case of CMB constraints
\citep[][]{ric08,ali17,pou17,ber17,nak18}, the injected energy density rate scales as $\dot{e}_{\rm
PBH} \propto f_{\rm PBH} \dot{M}^2_{\rm PBH}$, where $f_{\rm PBH}$ is the fraction of dark matter made
up of PBHs. Fixing $\dot{e}_{\rm PBH}$ and assuming for simplicity the same $M_{\rm PBH}$ and
$\dot{M}_{\rm PBH}$ for all PBH, a decrease of an order of magnitude in $\dot{M}_{\rm PBH}$ would allow an
increase of two orders of magnitude in $f_{\rm PBH}$. This would largely relax the present constraints
on $f_{\rm PBH}$ in the $10-100$~M$_\odot$ mass range. In addition to CMB constraints, constraints
derived from galactic emission \citep[][]{gag17,man19} could be also affected by mechanical feedback.

We have not discussed here the effects on the medium of the non-thermal radiation directly generated by the PBH outflows, because in our case outflows are rather weak, that is, $\epsilon\ll 1$. Notwithstanding, outflows from PBHs may have $\epsilon\approx 1$ and high radiative efficiency, in which case their high-energy radiation could contribute to heat, excite, and ionize the medium in addition to direct accretion emission. This would make the constraints on PBH abundance tighter, so such an effect should be considered in the future. In fact, radio \citep{gag17,man19} and high-energy emission from outflows produced by PBHs, located in dense regions of our galaxy, may be important enough to be detectable at galactic distances. This idea has been already explored in \cite{bar12b} in the case of isolated stellar BHs.


\section*{Acknowledgements}
We thank the anonymous referee for constructive and useful comments that helped to improve the manuscript.
We acknowledge N\'uria Torres-Alb\`a, Edgar Molina, Licia Verde, and Alexander James Mead for useful comments and discussions. 
V.B-R. acknowledges support by the Spanish Ministerio de Econom\'{i}a y Competitividad (MINECO/FEDER, UE) under grant AYA2016-76012-C3-1-P, with partial support by the European Regional Development Fund (ERDF/FEDER), MDM-2014-0369 of ICCUB (Unidad de Excelencia `Mar\'{i}a de Maeztu'), and the Catalan DEC grant 2017 SGR 643. NB is supported by the Spanish MINECO under grant BES-2015-073372.

\bibliographystyle{aa}
\bibliography{biblio}


\end{document}